# Metabolite patterns reveal regulatory responses to genetic perturbations

Tolutola Oyetunde, Jeffrey Czajka, Gang Wu, Cynthia Lo, and Yinjie Tang *

Department of Energy, Environmental and Chemical Engineering, Washington University, Saint Louis, Missouri, 63130.

*corresponding author.

## Abstract

Progress in metabolic engineering and synthetic biology for varied applications is strongly dependent upon detailed insights into cellular physiology and metabolism. Genetic and environmental perturbation experiments have been used to study microbes in a bid to gain insight into transcriptional regulation, adaptive evolution, and other cellular dynamics. These studies have potential in enabling rational strain design. Unfortunately, experimentally determined intracellular flux distributions are sometimes inconsistent or incomparable to each other due to different experimental conditions and methodologies.

Computational strain design relies on constraint-based reconstruction and analysis (COBRA) techniques to predict the effect of gene knockouts such as flux balance analysis (FBA), regulatory on/off minimization(ROOM), minimization of metabolic adjustment (MOMA), relative optimality in metabolic networks (RELATCH). Most of these knock-out prediction methods are based on conserving inherent flux patterns (between wild type and mutant) that are thought to be representative of the cellular regulatory structure. However, it has been recently demonstrated that these methods show poor agreement with experimental data.

To improve the fidelity of knockout predictions and subsequent computational strain design, we developed REMEP, a metabolite-centric method. We demonstrate the improved performance of REMEP by comparing the different methods on experimental knockout data of *E. coli* and *S. cerevisiae* grown in batch cultures. REMEP retains most of the features of earlier algorithms but is much more accurate in capturing cellular response to genetic perturbations. A primary reason for this is that REMEP relies on the assumption that cellular regulatory structure leaves a signature on metabolite patterns and not just flux patterns. REMEP will also prove useful in uncovering novel insights into cellular regulation and control.

# Introduction

Recent advances in metabolic engineering and synthetic biology have enabled cells to be used for a variety of industrial applications, including production of food and beverages, commodity chemicals, specialty chemicals, and pharmaceuticals. Metabolic engineers mainly use genetic approaches to adjust metabolic regulatory networks, and as such, are highly dependent on a detailed understanding of cellular regulatory and metabolic systems. These understanding must incorporate the role of that cellular events play in control of fluxes, such as feedback inhibition, structural modifications of enzymes, and enzyme synthesis[1]. Genetic and environmental perturbation experiments have been used to generate insights into these systems, including transcriptional regulation[2], adaptive evolution responses[3], and metabolic network robustness[4]. Additionally, the construction of the Keio library, which contains flux information on single gene KO *E. coli* mutants[5], is helping to guide these efforts. This information helps elucidate the structure of cellular regulation and how much control each enzyme/regulatory network exhibits on the metabolic flux/metabolic pool size. Unfortunately, much of the experimentally determined intracellular flux distributions are often inconsistent or incomparable with each other due to different experimental conditions and methodologies[6]. Despite the discrepancies, these quantitative studies of cellular responses provide relevant data that has the potential in enabling rational strain design.

Accurately predicting the metabolic flux redistribution of KO strains is vital for directing rational strain design. To facilitate these efforts, computational methods have been developed to predict cellular responses to genetic perturbations. The most prominent computational tools used are constraint-based reconstruction and analysis (COBRA) models. COBRA-based models require only the metabolic network stoichiometry and a defined 'objective function'[7] to predict cellular fluxes and have been used to guide metabolic engineering[8], drug discovery[9], and adaptive evolution studies[10]. A variety of COBRA models have been developed, including flux balance analysis (FBA)[11], minimizations of metabolic adjustment (MOMA)[11], regulatory on/off minimization of metabolic fluxes (ROOM)[12], and relatively optimality of in metabolic networks (RELATCH)[13]. FBA models use the assumption that microbes will display optimal growth during its prediction of metabolic flux distributions. MOMA and ROOM attempt to improve upon the FBA predictions by predicting a wild-type microbe's flux and using this predicted flux distribution as a reference state. The MOMA and ROOM algorithms then attempt to minimize the Euclidean and Hamming distances, respectively, between the mutant flux distribution and the wild-type[12]. The MOMA algorithm tends to favor many small changes in the mutant's metabolic flux network, while ROOM minimizes the number of significant changes. The RELATCH model uses experimental flux and expression data from a reference strain as a starting point. The algorithm first minimizes the number of regulatory changes in the mutant, and then activates previously latent pathways to for a final metabolic flux prediction.

Experimental verification of FBA predictions of mutant growth rates have been shown to only be representative of mutant strains that have undergone adaptive evolution and which are growing under optimal conditions[14,15]. MOMA, ROOM and RELATCH have all been shown to improve the prediction of mutant growth rates under certain conditions relative to FBA[11–13]. Despite the accurate growth rate predictions, there are often discrepancies in intracellular flux distributions, which may guide experimental efforts in the wrong direction[6]. These inaccuracies imply that the models do not accurately capture the method of cellular regulation and therefore, cannot provide novel insight into the understanding of cellular regulation. To improve the fidelity of knockout predictions and subsequent computational strain design, we developed REMEP, a metabolite-centric method. REMEP relies on the assumption that cellular regulatory structure leaves a signature on metabolite patterns and not just flux patterns. We demonstrate the improved performance of REMEP by comparing the different methods on experimental knockout data of *E. coli* and *S. cerevisiae* grown in batch cultures. More importantly, unlike earlier methods, REMEP, allows for the utilization of multiple knock-out experiments (as well as other model predictions) to improve prediction fidelity cumulatively and systematically. REMEP will also prove useful in uncovering novel insights into cellular regulation and control.

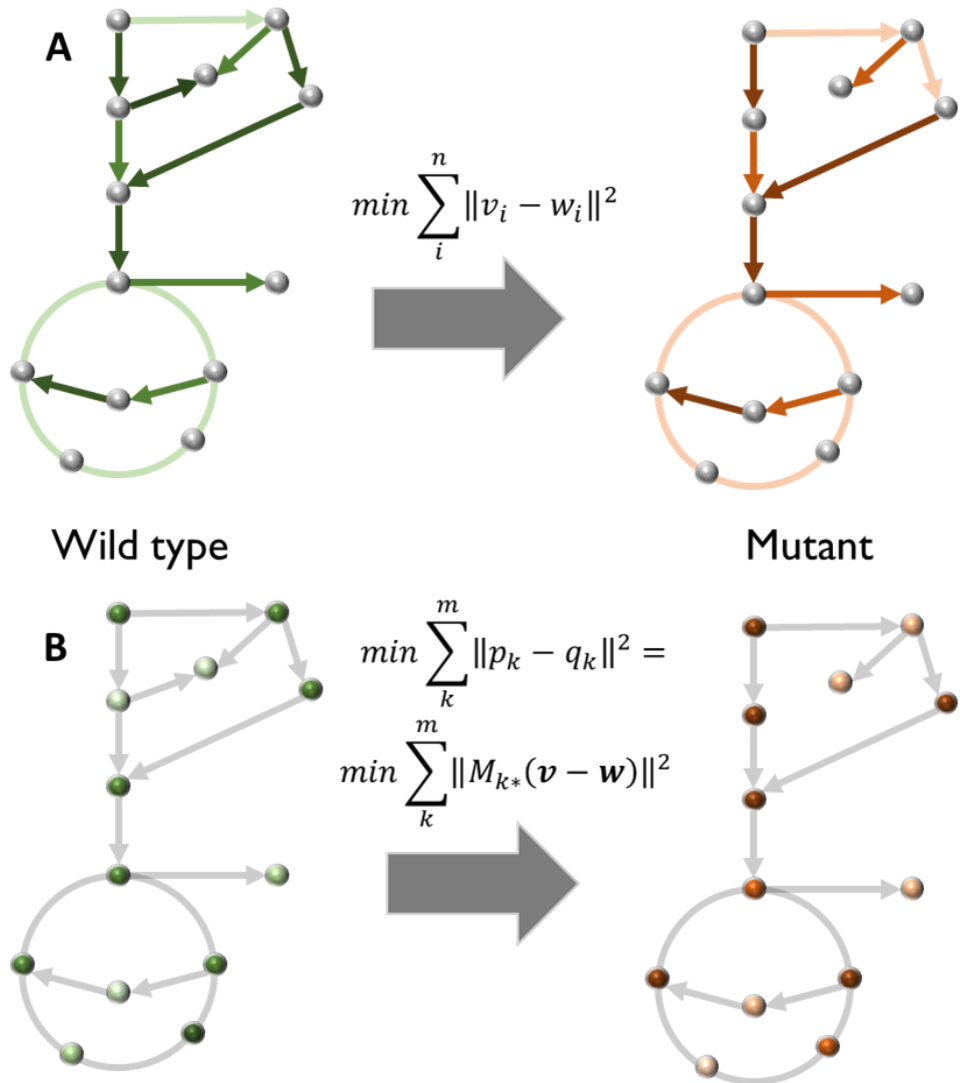

**Figure 1: Comparison of REMEP to earlier mutant prediction algorithms.** A) MOMA, RELATCH and ROOM abstract cellular regulation as an attempt to conserve flux patterns. MOMA minimizes the Euclidean norm between wild type and mutant flux distributions, ROOM minimizes the number of larges changes in flux distribution and RELATCH minimizes flux distributions with an additional constraint/objective function term based on gene expression B) REMEP hypothesizes metabolite patterns (which are essentially formulated as weighted flux patterns) as a manifestation of basic cellular regulation. n by 1 vectors ***v*** and ***w*** represent wild-type and mutant fluxes respectively while m by 1 vectors p and q represent the fluxes through metabolites in the wild type and mutant strains. Matrix ***M*** is a m by n 'weighting' matrix formulated from the stoichiometric matrix (described in the methodology section). Arrows represent the fluxes through the enzyme, circles represent the metabolite pool, and the colors represent the conserved portion in each method. Darker shades represent larger fluxes.

## Results and Discussion

We compared the predictions of REMEP to existing algorithms (FBA, MOMA and RELATCH) using knockout datasets of *E. coli* and *S. cerevisiae* strains. Genome-scale models of *E. coli* (iAF1260) and S. cerevisiae (iMM904) were downloaded from the BiGG database[16]. iAF1260 was used for *E. coli* because no significant difference was observed with other *E. coli* models (iJO1366 and iJR904)[13]. The gene expression data for *E. coli*[17] and *S. cerevisiae*[18] needed for RELATCH computations was obtained from previously published work. 26 mutant flux distributions were obtained for *E. coli* (four from 19 s[19] and 22 from 7. s[7]). 36 *S. cerevisiae* mutant strain flux distributions were obtained from [20]. All simulations were run in MATLAB. The COBRA toolbox implementations of FBA and MOMA were used to obtain predictions for the models, while the RELATCH code was downloaded from http://reedlab.che.wisc.edu/

### A) *E. coli* Mutants

The flux data for four single gene knockouts in *E. coli* was reported in (reference 19). Figure 2 shows the comparison of the different algorithms' flux predictions for each mutant. REMEP shows consistently high correlation with experimental data. The REMEP prediction is better than the FBA and MOMA models, and on par with the RELATCH predictions. REMEP does not use gene expression data unlike RELATCH.

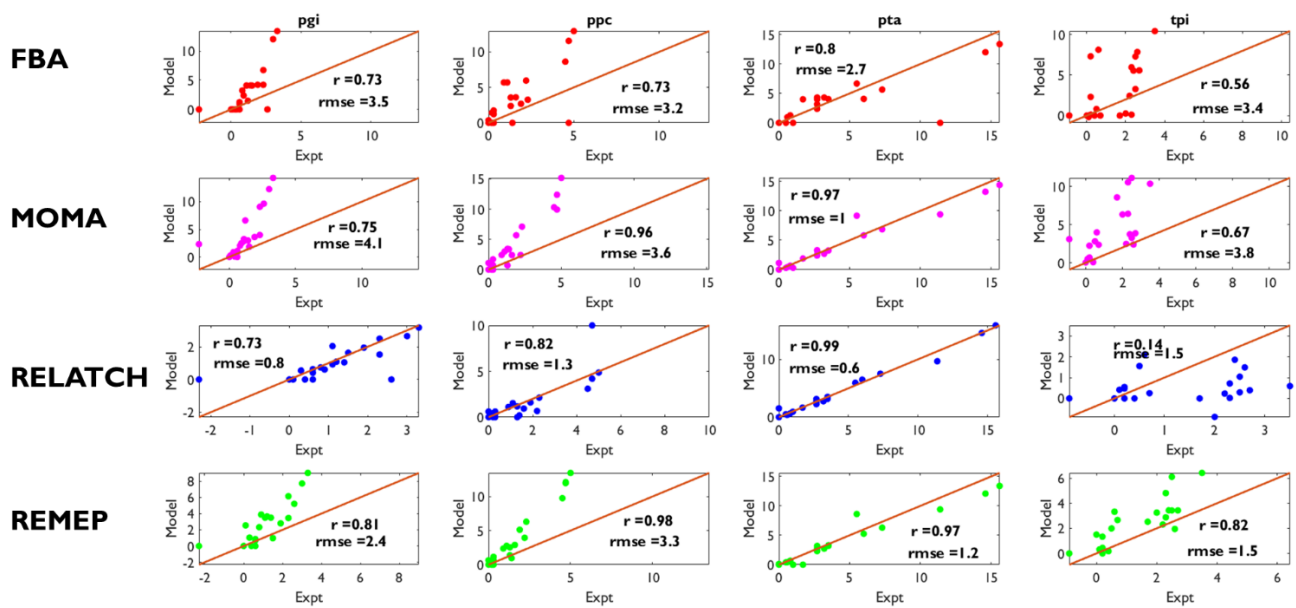

**Figure 2. Comparison of different mutant phenotype prediction algorithms on four *E. coli* mutant strains** (genes knocked out: pgi, ppc, pta and tpi). r is the pearson's correlation coefficient while rmse is the root mean square error. Points correspond to the experimentally measured fluxes in central metabolism.

The flux data for 22 *E. coli* mutants grown ion glucose in a batch reactor was reported in (reference 7). In Figure 3, we compare the root mean square errors of the experimental data and model predictions for biomass growth, glucose uptake and acetate secretion rates for these mutants. REMEP achieves reasonable accuracy for both the internal and external fluxes (Supplementary figure (as in a rmse/r comparison as well, maybe just phenotypes??).

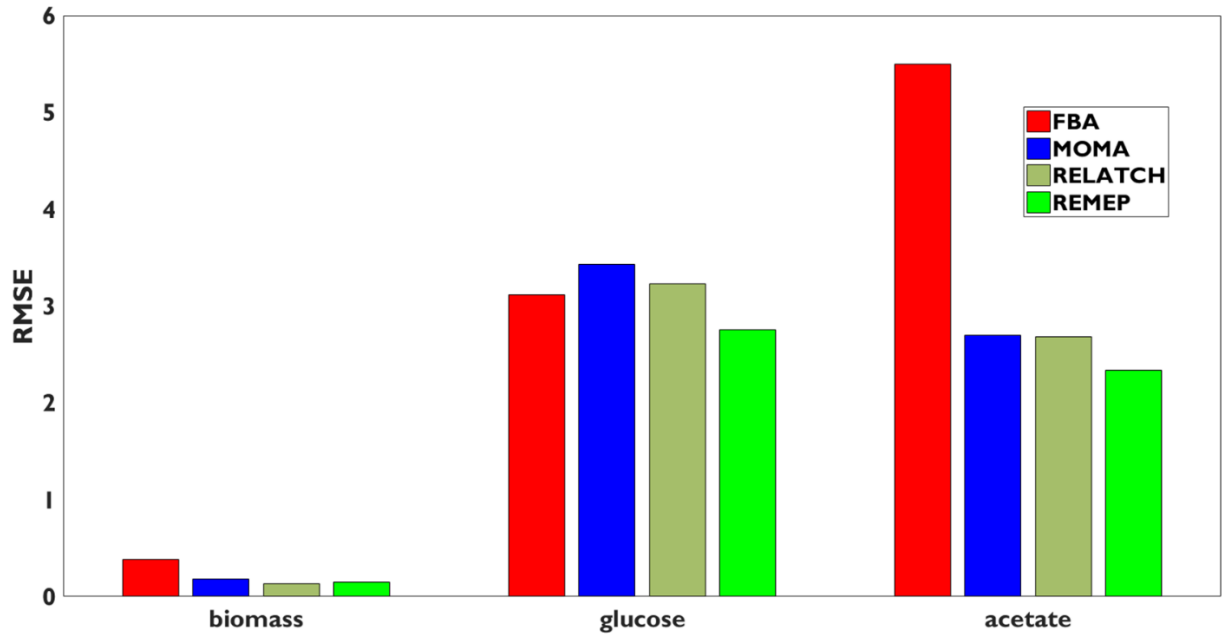

**Figure 3. Root mean square error between model predictions and experiments for measured external fluxes and biomass growth of 22 *E. coli* knockouts.**

**B) *S. cerevisiae* Mutants**

Figure 4 shows the comparison of the RELATCH and REMEP algorithms' flux predictions for 8 of single gene knock-outs in *S. cerevisiae*[18]. REMEP model predictions are better or on par with RELATCH and outperform both FBA and MOMA algorithms as shown by figure 5 and Supplementary Figure XXX). By focusing on metabolite patterns instead of flux patterns, REMEP captures subtleties in cellular regulation that are not possible with the earlier methods, which are all based on some form of conservation of flux patterns between mutant and wild type strains. A good example is the *E. coli* tpi mutant (Figure 2) predictions in which REMEP vastly outperforms the other algorithms.

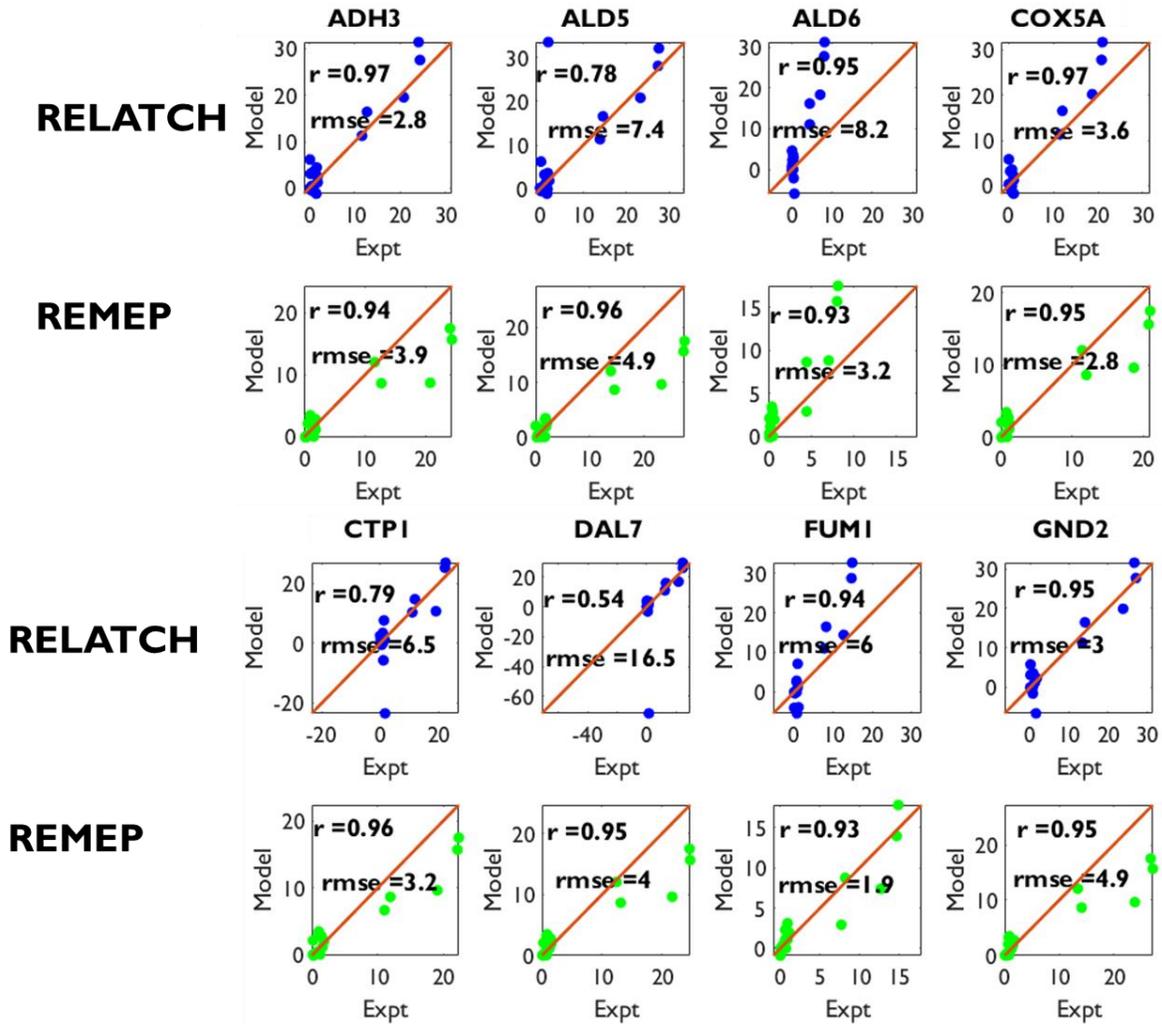

**Figure 4. Comparison of RELATCH and REMEP on eight *S. cerevisiae* mutant strains.** r is the Pearson's correlation coefficient while rmse is the root mean square error. The full set of 36 mutant strains for each of the four prediction algorithms studied (FBA, ROOM, RELATCH and REMEP) is presented in the supplementary file

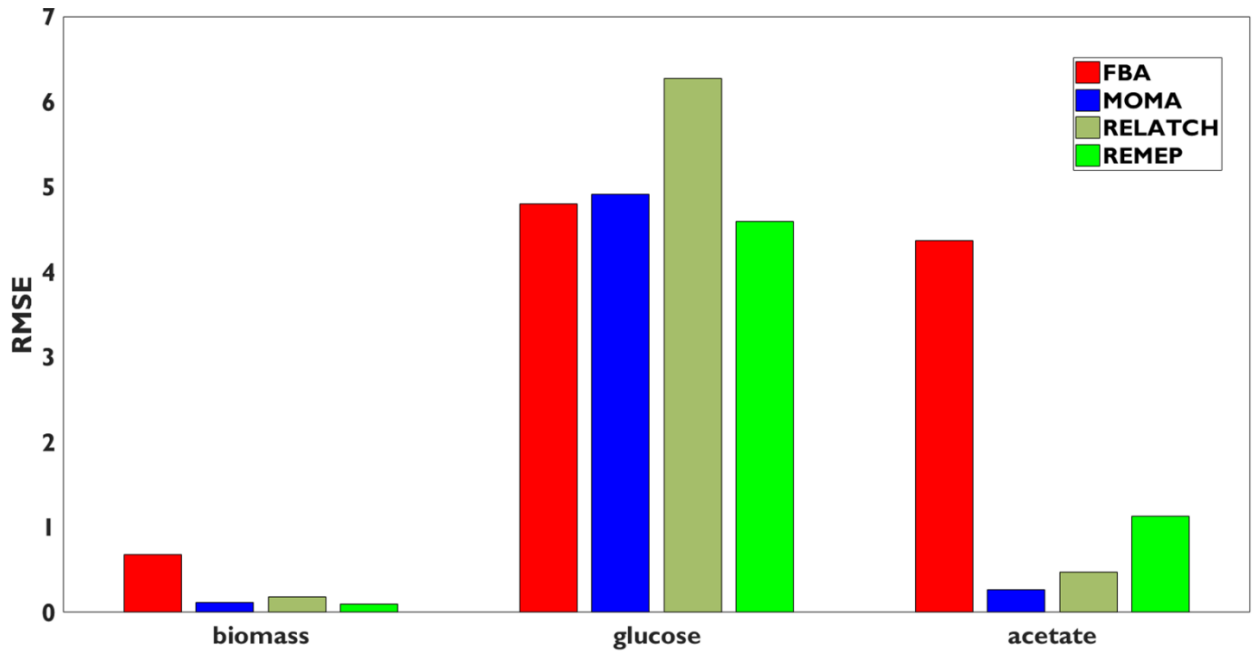

**Figure 5. Root mean square error between model predictions and experiments for measured external fluxes and biomass growth of 36** *S. cerevisiae* **knockouts.**

## C) Cellular regulatory structure implied by different mutant prediction algorithms

As many computational strain design tools rely on mutant prediction algorithms, it is important to have a mutant prediction algorithm that accurately reflects the cellular regulatory structure. REMEP aims to fulfil that objective by capturing cellular regulatory behavior encoded in fluxes through metabolite nodes and patterns, which have been shown to contain useful information about cellular function and evolutionary trends[21]. A recent study successful demonstrated the utility of metabolite patterns to the classic problem of gap filling of genome-scale metabolic network reconstructions[22].

In this work, we highlight the fact that the hypothesis made by different mutant prediction algorithms implies a cellular regulatory structure pattern that can be studied. This is demonstrated in Figure 6 where we show the percentage change in the usage of selected reactions in central metabolism of *E. coli* and *S. cerevisiae* after genetic knockout. A key difference between *E. coli* and *S. cerevisiae* is that there are more significant changes in *E. coli* flux distribution upon genetic modification. Thus, the cellular regulatory structure of smaller prokaryotes is predicted to be more sensitive to gene knockouts than eukaryotes. We also note the similarity between RELACTH and REMEP *E. coli* patterns even though REMEP does not make use of gene expression data. This suggest that most of the information embedded in gene expression can be observed from the metabolite patterns. Thus, REMEP serves as a useful substitute for RELATCH when gene expression data is not available and can easily be incorporated into computational strain design tools[23–27]. Comparison of heat maps shown in Figure 6 with experimentally generated ones can help pinpoint areas of improvement and refinement for mutant prediction tools.

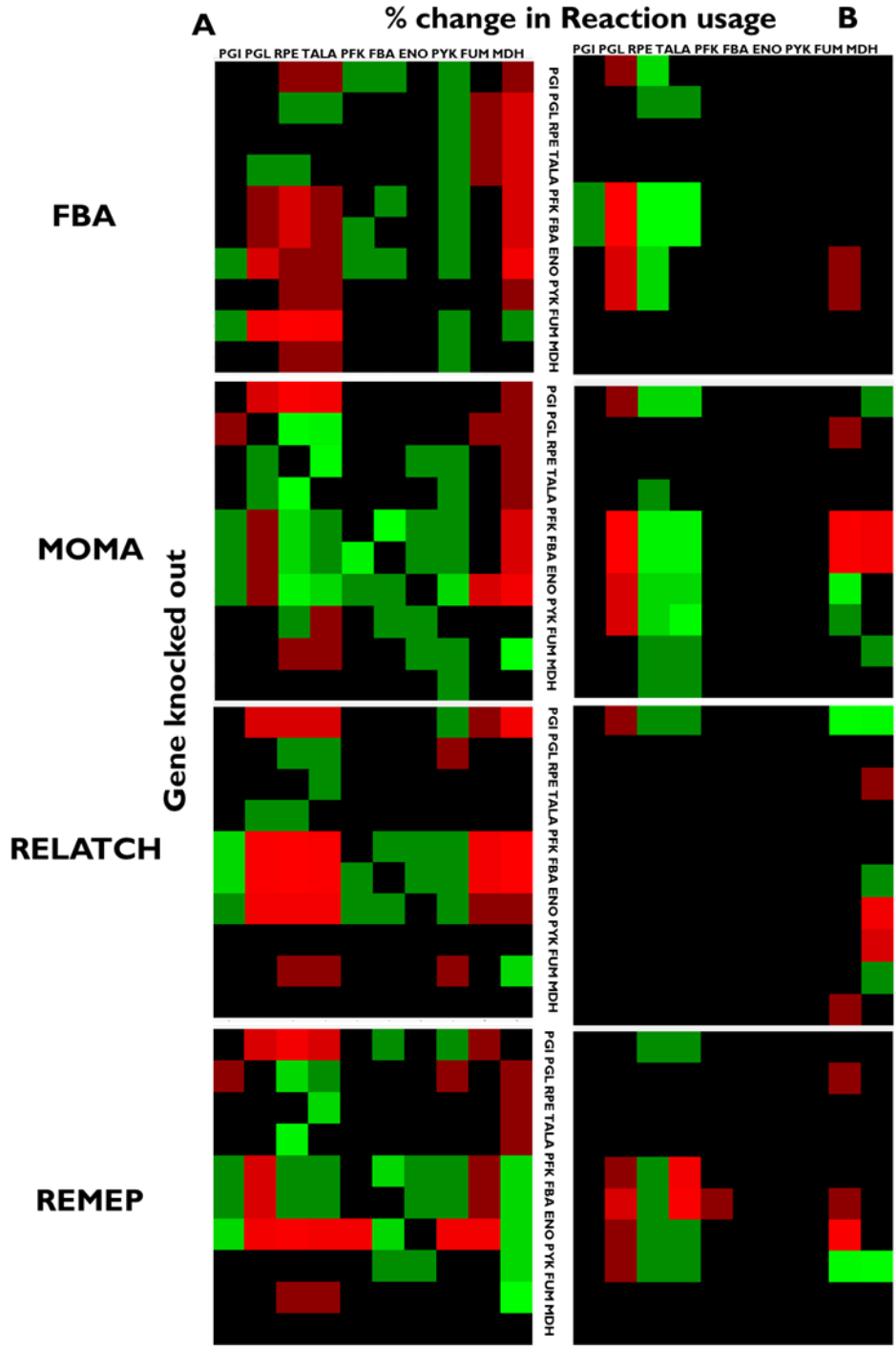

**Figure 6. Heat Map showing percentage change in selected reactions of central metabolism of A) E. coli and B) S. cerevisiae upon gene knockout.** For each simulation, all the genes associated with the reaction were silenced. Full maps with actual percentage numbers are provided in the supplementary files.

# Methodology

## Mathematical Formulation of REMEP

Consider a $m$ by $n$ stoichiometric matrix $S$ representing the metabolism of an organism with $m$ metabolites and $n$ reactions such that at steady state the following equation is fulfilled:

$$S.w = 0 \qquad (1)$$

Where $w$ is the vector of reactions (fluxes), reversible and irreversible. We can rewrite each reversible flux in $w$ as the difference between two irreversible fluxes and expand S accordingly so we have:

$$S^*.w^* = 0 \qquad (2)$$

Where $S^*$ is m by $n+r$ matrix and $w^*$ is $n+r$ vector, $r$ being the number of reversible reactions.

Furthermore, for each metabolite $i$, we can write a vector $M_i$ consisting of only the positive elements in the row $i$ of $S^*$ (that is, reactions producing the metabolite). We could thus construct a matrix $M$, such that

$$M.w^* = d \qquad (3)$$

Where each element in vector $d$ represents the total amount of producing flux through a metabolite.

REMEP minimizes the difference between the total flux through each metabolite for mutant and wild type strains by solving the following constrained least squares optimization problem:

$$min\|M.v^* - d\|^2$$

Subject to:

$$S^*.v^* = 0 \qquad (4)$$

$$0 \leq v^* \leq ub$$

scaled versions of the objective function could also be used such as:

$$min \left\| \frac{M.v^*}{\sum M.v^*} - \frac{d}{\sum d} \right\|^2 \qquad (5)$$

Minimization of the difference between biomass growth of wild type and mutant strains could also be added as an extra row in *M*. The values in the upper bound vector *ub* can be set based on experimental information. For example, if a reaction was knocked out, the corresponding element in *ub* would be set to zero. Note (4) could also be formulated as a quadratic optimization problem.

## Conclusions

We have presented a simple yet powerful mutant flux prediction algorithm, REMEP based on metabolite patterns captured by sum of fluxes through metabolite nodes. REMEP gives better prediction of regulatory and metabolic response to genetic perturbation in both prokaryote and eukaryote strains than comparable algorithms. REMEP can serve as a useful substitute to more complex prediction algorithms like RELATCH that rely on gene expression data. Metabolic regulation appears to conserve relative metabolite associations. REMEP will prove a useful for computational strain design tool for metabolic engineering as well as provide a platform for understanding the basis of cellular regulation.